\documentclass[bibyear]{aa}
\usepackage[varg]{txfonts}
\usepackage{graphicx}

\begin{document}

\title{Astrometric asteroid masses: a simultaneous determination}
\titlerunning{Simultaneous determination of asteroid masses}
\author{Edwin Goffin}
\authorrunning{Edwin Goffin}
\institute{Aartselaarstraat 14, B-2660 Hoboken, Belgium \\
\email{edwin.goffin@skynet.be}}
\date{Received 27 September 2013 / Accepted ... 2014}

\abstract
{Using over 89 million astrometric observations of 349\,737 numbered
minor planets, an attempt was made to determine the masses of 230
of them by simultaneously solving for corrections to all orbital elements
and the masses.
For 132 minor planets an acceptable result was obtained, 50 of which appear
to be new.}
\keywords{astrometry - celestial mechanics - minor planets, asteroids: general}
\maketitle

\section{Introduction}
The astrometric determination of asteroid masses is the technique of
calculating the mass of asteroids (hereafter called ''perturbing asteroids''
or ''perturbers'') from the gravitational perturbations they exert
on the motion of other asteroids (called ''test asteroids'').

The usual procedure to calculate the mass of one asteroid is, first,
to identify potential test asteroids by searching for close approaches
with the perturbing asteroids and then computing some parameter
(usually the deflection angle or the change in mean motion) characterising
the effect of the perturbing asteroid, see e.g. Gal\'ad and Gray~(\cite{Gala2002}).
Promising cases are then further investigated to see if they yield meaningful
results, since the outcome also depends on the number, distribution and
quality of the available astrometric observations of the test asteroid.
Asteroids not making close approaches but moving in near-commensurable orbits
can also be useful for mass determinations (Kochetova~\cite{Koch2004}).

A better but more laborious way to find suitable close encounters is to 
integrate the orbits of test asteroids over the period of available observations
with and without the perturbations of the perturbing asteroid (Michalak~\cite{Mich2000}).
Significant differences in the geocentric coordinates between the two calculations
indicate cases that should be investigated further.

Initially, asteroid masses were computed from their perturbations on one other
asteroid, the first instance being the mass of 4~Vesta computed from the orbit
of 197~Arete by Hertz~(\cite{Hert1968}).
When several mass values for the same asteroid are calculated from a number of
test asteroids, the usual practice is to compute a weighted mean along with
the associated uncertainty as the final value, as in Michalak~(\cite{Mich2000}).
Mass determinations where one or a few masses are determined simultaneously
from a number of test asteroids have been attempted on a modest scale in
the past--see Sitarski and Todororovic-Juchniewicz~(\cite{Sita1995}),
Goffin~(\cite{Goff2001}) and others.

This paper gives the results of an attempt to simultaneously determine a
large number of asteroid masses (230) from a substantial number of test
asteroids (349\,737).
Sect. 2 explains the basic principles and problems of simultaneous asteroid mass
determination and in Sect. 3 the details of the actual calculations are given.
Sect. 4 briefly deals with the observations used and explains the statistical
analysis of the residuals that was used to assign weights to, or reject observations.
In Sect. 5 the results are presented and Sect.6 discusses them.

Throughout this paper, asteroid masses are expressed in units of $10^{-10}$
solar masses ($M_{\sun}$) and generally given to 3 decimal places.

\section{Simultaneous mass determination}
The principle of the simultaneous determination of asteroid masses is simple:
use the astrometric observations of $N$ test asteroids to determine amongst others
the masses of $M$ perturbing asteroids.
This direct approach has a number of distinct advantages.
First, there is no need to search for close encounters to the $M$ perturbers,
assess their efficiency and select promising pairs for the actual calculations
(the same can be said for near-commensurable orbits).
All test asteroids are treated equal and close approaches will appear anyway
when numerically integrating the $N$ orbits.
Next, one doesn't have to compute weighted means of masses and their standard deviations.
Also, the mutual correlations between the $M$ masses are taken into account
in the global solution.
But all this comes at a cost: computer resources!

\subsection{The normal equations}
In the process of mass determination, the equations of condition of the
linearised problem are usually transformed into the normal equations :
\begin{equation} \label{eq01}
  \vec{A} \cdot \vec{x} = \vec{b}
\end{equation}
The coefficient matrix $\vec{A}$ is square, positive definite and symmetric
about the main diagonal.
For an ''ordinary'' mass determination ($N = 1,M = 1$) it is of size
$7~\times~7$ (rows~$\times$~columns).
Note that the coefficients below the main diagonal do not need to be stored,
but that part of the matrix is often used as temporary storage when solving
the equations or computing the inverse $\vec{A}^{-1}$.

The normal equations for simultaneous mass determination ($N > 1,M \geq 1$) are
easily assembled from the normal equations of the individual test asteroids.
Sitarski and Todororovic-Juchniewicz~(\cite{Sita1992}) give a schematic representation
for a simple case ($N = 2,M = 1$).
In general, $\vec{A}$ is of size $(6N + M) \times (6N + M)$.
While this is not prohibitive for modest applications, it is clear that $\vec{A}$
will grow rapidly in size with large values of $M$ and especially $N$,
and eventually exceed the available computer memory.

The full normal equations (\ref{eq01}) can be written as:
\begin{equation} \label{eq06}
 \left|
 \begin{array}{c@{\hspace{0.5em}}c@{\hspace{0.5em}}c@{\hspace{0.5em}}c@{\hspace{0.5em}}c@{\hspace{0.5em}}c}
   A_{1,1}   &        0  &      0    & \cdots &      0    & A_{1,M} \\
        0    & A_{2,2}   &      0    & \cdots &      0    & A_{2,M} \\
        0    &        0  & A_{3,3}   & \cdots &      0    & A_{3,M} \\
    \cdot    &    \cdot  &  \cdot    & \cdots &      0    & \cdot   \\
    \cdot    &    \cdot  &  \cdot    & \cdots &      0    & \cdot   \\
        0    &        0  &      0    & \cdots & A_{N,N}   & A_{N,M} \\
   A_{1,M}^T & A_{2,M}^T & A_{3,M}^T & \cdots & A_{N,M}^T & A_{M,M}
 \end{array}
 \right|
 \left|
 \begin{array}{c}
   X_1   \\
   X_2   \\
   X_3   \\
   \cdot \\
   \cdot \\
   X_N   \\
   X_M
 \end{array}
 \right|
 =
 \left|
 \begin{array}{c}
   B_1   \\
   B_2   \\
   B_3   \\
   \cdot \\
   \cdot \\
   B_N   \\
   B_M
 \end{array}
 \right|
\end{equation}
where:
\begin{itemize}
 \item
   $A_{1,1}, \dots, A_{N,N}$ are the $6 \times 6$ blocks of the partial derivatives for
   the 6 orbital elements of the $N$ test asteroids
 \item
   $A_{1,M}, \dots, A_{N,M}$ (each of size $6 \times M$) and $A_{M,M}$ (size $M \times M$)
   correspond to the $M$ masses
 \item
   $X_1, \dots, X_N$ are the variables pertaining to the $N$ asteroids, namely
   the corrections to the 6 orbital elements (or components of the state vector) at epoch
 \item
   $X_M$ are the variables pertaining to all perturbing asteroids,
   namely the corrections to the $M$ masses.
\end{itemize}
The zeros represent of course $6 \times 6$ blocks of zeros.
The matrix clearly contains a staggering amount of elements filled with zeros!

\subsection{Solving the normal equations}
One way to reduce computer storage is to store only the non-zero blocks of \vec{A}:
\begin{equation} \label{eq11}
 \vec{A} = \left|
 \begin{array}{cc}
   A_{1,1}   & A_{1,M}  \\
   A_{2,2}   & A_{2,M}  \\
   A_{3,3}   & A_{3,M}  \\
   \cdot     & \cdot    \\
   \cdot     & \cdot    \\
   A_{N,N}   &  A_{N,M} \\
             &  A_{M,M}
 \end{array}
 \right|
\end{equation}
When stored as a single array, the size of $\vec{A}$ is thus reduced
to $(6N + M) \times (6 + M)$, or, if $A_{M,M}$ is stored separately,
to $6N(M + 6) + M\times M$.
The Cholesky decomposition algorithm used to invert matrix $\vec{A}$ was
adapted to cope with this new form.
In that way, long test calculations were done up to $N = 300\,000, M = 105$ (this
proved to be a practical upper limit).
Then, following a suggestion by E. Myles Standish, a formulation was
adopted that only requires a matrix of dimensions $(6 + M) \times (6 + M)$!

The normal equations (\ref{eq06}) can be written as the pair:
\begin{equation} \label{eq21}
 \left\lbrace
 \begin{array}{rcl}
   A_{i,i} X_i + A_{i,M} X_M &=& B_i, \hspace{1em}i=1,N	\\
   \sum_i(A_{i,M}^T X_i) + A_{M,M} X_M &=& B_M
 \end{array}
 \right.
\end{equation}

The first step is to eliminate the $X_i$'s from the second equation.
From the first, we get:
\begin{equation} \label{eq26}
  X_i = A_{i,i}^{-1} (B_i - A_{i,M}X_M)
\end{equation}
Substituting this in the second and rearranging gives:
\begin{equation} \label{eq31}
  X_M = \left[A_{M,M} - \sum_i\left(A_{i,M}^T A_{i,i}^{-1} A_{i,M}\right)\right]^{-1}
		\left[B_M - \sum_i\left(A_{i,M}^T A_{i,i}^{-1} B_i\right)\right]
\end{equation}
With the $X_M$'s one can now solve eq. (\ref{eq26}) for the $X_i$'s.

\subsection{Standard deviations and correlations}
The formal standard deviations and correlations come from the inverse of the
\vec{A}-matrix.
Representing the components of $\vec{A}^{-1}$ by $\alpha$'s, we have from
the definition:
\begin{equation} \label{eq36}
 \left|
 \begin{array}{cc}
  A_{i,i}   & A_{i,M} \\
  A_{i,M}^T & A_{M,M}
 \end{array}
 \right|
 \left|
 \begin{array}{cc}
  \alpha_{i,i}   & \alpha_{i,M} \\
  \alpha_{i,M}^T & \alpha_{M,M}
 \end{array}
 \right|
 = I
\end{equation}
The multiplication gives:
\begin{equation} \label{eq41}
 \begin{array}{lclcl}
   A_{i,i} \alpha_{i,i}   & + & A_{i,M} \alpha_{i,M}^T & = & I	\\
   A_{i,i} \alpha_{i,M}   & + & A_{i,M} \alpha_{M,M}   & = & 0	\\
   A_{i,M}^T \alpha_{i,i} & + & A_{M,M} \alpha_{i,M}^T & = & 0	\\
   A_{i,M}^T \alpha_{i,M} & + & A_{M,M} \alpha_{M,M}   & = & I
 \end{array}
\end{equation}

The second equation of (\ref{eq41}) gives:
\begin{equation} \label{eq46}
  \alpha_{i,M} = -A_{i,i}^{-1} A_{i,M} \alpha_{M,M}
\end{equation}
Substituting this in the fourth yields:
\begin{equation} \label{eq51}
  \alpha_{M,M} = \left[A_{M,M} - \sum_i\left(A_{i,M}^T A_{i,i}^{-1} A_{i,M}\right)\right]^{-1}
\end{equation}
which has already been calculated and inverted in (\ref{eq31}).

With (\ref{eq46}) one gets $\alpha_{i,i}$ from the first equation of (\ref{eq41}):
\begin{equation} \label{eq56}
  \alpha_{i,i} = -A_{i,i}^{-1} (I - A_{i,M} \alpha_{i,M}^T)
\end{equation}
The sequence of the computation is thus: (\ref{eq51}), (\ref{eq46}), (\ref{eq56}).

\subsection{Test asteroids}
The test asteroids are those among the first 350\,000 numbered asteroids with
perihelion distance smaller than 6.0 au (to exclude distant objects
such as KBO's \dots), a total of 349\,737.
Their orbits were improved using perturbations by the major planets,
Ceres, Pallas and Vesta, before using their orbital elements and observations
as input for the simultaneous mass determination.

The positions of the eight major planets and the Moon are taken from the
JPL~ephemeris DE422, which is an extended version of DE421 (Folkner et al.~\cite{Folk2008}).
The coordinates of the three largest asteroids initially come from a separate
orbit improvement of these bodies.

\subsection{Perturbing asteroids}
Selecting the perturbing asteroids is basically quite simple: just take those
with the largest mass. Reliable masses have been calculated for a number
of asteroids, but for most others one must rely on estimates based on the
diameter and the taxonomic class.
Mass estimates for these asteroids were computed assuming a bulk density of
1.80~$\mathrm{g/cm^3}$ for class C (including B, D, F, G, P, T and X),
2.20~$\mathrm{g/cm^3}$ for class S (including A, E, K, Q, R and V) and
4.20~$\mathrm{g/cm^3}$ for class M.

Diameters were selected from the sources given in Baer et al.~(\cite{Baer2011}),
or from occultation results (Broughton~\cite{Brou2011}),
or else from the results of the WISE (Masieri et al.~\cite{Masi2011}) or
IRAS (Tedesco et al.~\cite{Tede1992}) surveys.
For 4~Vesta the value from the Dawn spacecraft was used (Russell et al.~\cite{Russ2012}).
The diameter of 52~Europa is from Merline et al.~(\cite{Merl2013}) and that of
349~Dembowska from Majaess et al.~(\cite{Maja2008}).
In case the dimensions of a tri-axial ellipsoid were given, the equivalent diameter
(of a sphere of the same volume) was calculated.
When none of these sources gave a diameter, a nominal value $D$ was computed from
the absolute magnitude $H$:
\[ log D  =  3.62  -  0.20H \]

In this way, mass estimates were calculated for the first 100\,000 numbered
minor planets and those with mass $> 0.005 \times 10^{-10} M_{\sun}$ and perihelion distance
$q < 6.0$ au, a total of 257, were withheld as perturbers for a first
test run with the 349\,737 test asteroids.
After a few iterations it turned out that 279~Thule and the Trojans in
the list didn't yield acceptable masses, so a new list was created with $q < 4.0$
and these 230 asteroids were used in the final calculations.

Though it would be possible to include the masses of the major planets in the solution,
they were held fixed at the DE422 values.

\section{Computations}
The basics of the computations have been given in many previous papers,
amongst others in Goffin~(\cite{Goff1991}) and Goffin~(\cite{Goff2001}).
The force model includes the Newtonian attraction due to the Sun,
the 8 major planets and the Moon, and the $M$ perturbing asteroids (one
less when the test asteroid is one of the perturbers).
The heliocentric coordinates of the major planets and the Moon are stored in a
planetary file together with those of the $M$ perturbing asteroids.
Also taken into account are the first-order relativistic effects due to the Sun
as given in Ivantsov~(\cite{Ivan2008}) and the effects of the solar oblateness.

The longest part of the calculations is the numerical integration of the
$N$ test asteroid orbits and of the differential equations of all the
required partial derivatives.
These calculations only became feasible with the advent of personal computers with
multi-core processors.

By the very nature of the problem, the solution is obtained by successive approximations.
Each iteration cycle consists of four consecutive steps, each actually corresponding
to a computer programme:
\begin{itemize}
 \item	\emph{P1}: Numerical integration of the $N$ test asteroids
 \item	\emph{P2}: Solution of the normal equations
 \item	\emph{P3}: Recomputation of the coordinates of the perturbing asteroids
 \item	\emph{P4}: Statistical analysis of the residuals.
\end{itemize}

The programme \emph{P1} takes care of the numerical integration of the $N$ orbits.
The mutual distances between perturbing and test asteroids are monitored and
the step size is decreased when necessary.
Orbital elements of all asteroids and masses of the perturbing asteroids
are read in from the elements file on disk.
The observations are weighted using the results of the statistical analysis,
after correcting for any magnitude effects and systematic bias (see Sect. 4).
The normal equations for each test asteroid are output to a disk file and
all residuals $(O-C)$ are stored in the residuals file together with relevant
information (observatory code, time, visual magnitude of the asteroid, etc.).

Programme \emph{P2} reads in the separate normal equations, assembles the normal
equations of the global solution, inverts the coefficient matrices and
computes the $6N + M$ unknowns (initially 2\,098\,652 in total),
their standard deviations and the correlation coefficients.
This is actually done twice during each iteration.
After the first solution, the validity of the resulting $M$ masses is checked.
Following Baer and Chesley~(\cite{Baer2008a}), a mass is considered acceptable
when its \emph{significance}, this being the ratio of the mass to its standard
deviation, is larger than 2 ($s_i = m_i/\sigma_{m_i} > +2.0$), \emph{and} if the
resulting density lies between 0.5 and 8.0 $\mathrm{g/cm^3}$.
The rows and columns of the normal matrix corresponding to unacceptable
masses are then modified (by filling in 0's or 1), and the right-hand side
is changed so that the second solution will force them back to their
initial (estimated) values.
The improved masses and orbital elements from this solution are stored in
the elements file.

The third programme (\emph{P3}) uses the improved elements of the $M$ perturbing
asteroids to compute new heliocentric rectangular coordinates and copies them
into the planetary file.

Programme \emph{P4} uses the residuals file to perform a statistical analysis of the
performance of all observatories; the detailed description is the subject of Sect. 4.
The results are stored on disk and used by \emph{P1} to assign weights to
the observations. This analysis is done every 3 iterations.
The sequence of the computations is thus: $3 \times (P1+P2+P3) + P4$.

\section{Observations, weights and outlier rejection}
\subsection{Astrometric observations}
The principal source of minor-planet observations are the files maintained at
the Minor Planet Center (Cambridge, Mass.); the version of January 2013 was used.
Throughout the years I have re-reduced most of the older observations of
the first 1000 numbered asteroids made since their discovery and published
in the open literature.
They consist of series of visual micrometer and meridian observations, and a few
series of photographic observations.
Wherever possible, they were re-reduced with the aid of modern star catalogues.
No further details will be given here.
These observations were added to -- and replaced some of -- the data from the MPC.
For the optical observations, those given to a precision worse than
0.01s in right ascension or 0.1\arcsec in declination were not used.
The total number of observations of the $N$ test asteroids is over 89.4 million.

In the past, most researchers have used various methods for rejecting outliers
and assigning weights to the observations.
Some have used an iterative $n\sigma$ rejection criterion, or have assigned
a-priori weights depending on the type of observation and/or the time period.
Such simplifications may be easy to implement but do not reflect the
real accuracy of the observations.
Relevant weights must therefore be derived from a statistical analysis of the
residuals resulting from orbit improvements.
Since these two steps are interdependent, this is an iterative procedure.

Serious thoughts to the subject were first given by M.~Carpino
	\footnote{Then on his web site; not available any more now.}
and a thorough analysis was presented in Carpino et al.~(\cite{Carp03}).
However, at that time I decided to continue my own statistical analysis for two reasons.
The Carpino et al. paper was not yet published when I embarked on the
project of mass determinations, so no practical weighting scheme was available.
Also, the many old observations which I had re-reduced are mostly not present
in the MPC data set and thus I had to provide their statistics myself.

\subsection{Grouping of observations}
The complete set of observations must be divided into groups with (supposedly)
homogeneous properties, i.e. by instrument (observatory, telescope),
observation and reduction method, and further by any type of information that
can be relevant (observer, type of measuring instrument, number of readings taken etc.).
In practice, all this information is not readily available and has not been
stored with the current observation format of the Minor Planet Center.
Therefore a division was made per observatory code (which can thus refer to
different instruments and observers) and further by the following 4 observational
methods: transit and meridian (T), visual micrometer (M), photographic (P) and CCD (C).

Each of these observation groups was further split up into a number of time bins
as follows.
A first time bin was "filled" until its time length exceeded 2500 days \emph{or}
the total number of observations exceeded 100\,000; then the next bin was begun.
The first condition limits the time length of the bins for observatories with few
observations, while the second allows for a more detailed analysis of the more
productive surveys.
Time bins with less than 50 observations were considered too small
for analysis and were combined into the 4 groups of observatory code 500~Geocentric.
This code groups all observations that have been corrected for parallax
and the reference to the original observatory is thus lost anyway.

\subsection{Statistical analysis}
One analysis cycle consists of the following steps.
\begin{itemize}
 \item
  Using the residuals file created by \emph{P1}, the observations are grouped
  in time bins as decribed above.
\begin{figure}
  \resizebox{\hsize}{!}{\includegraphics{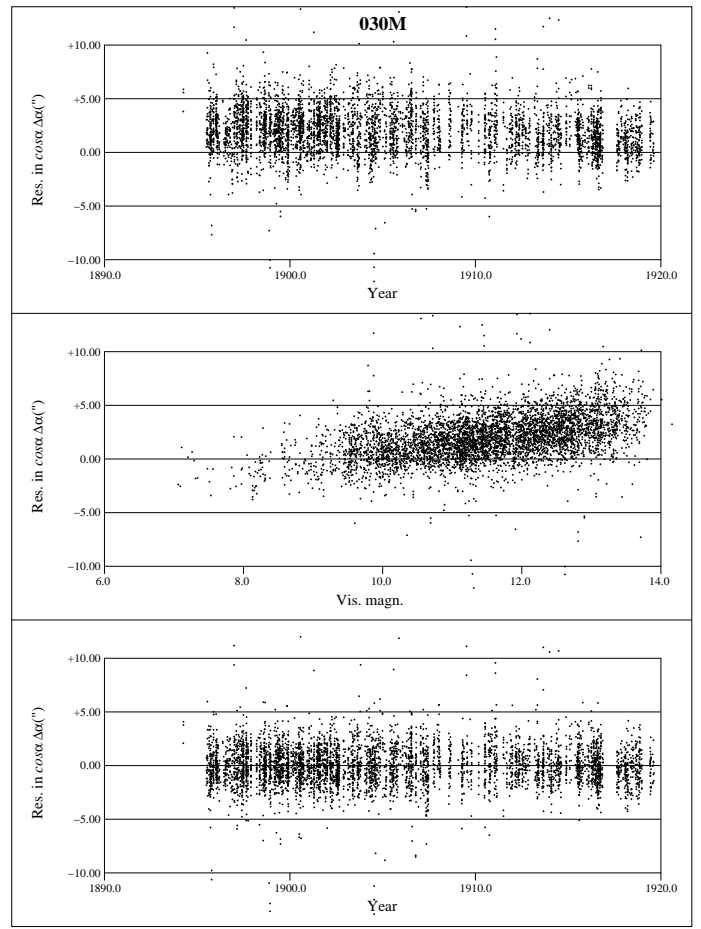}}
  \caption{Magnitude equation in right ascension for the visual micrometer
	observations made at 030~Arcetri (5147 observations, 1894--1919).
	Top graph: residuals versus time, uncorrected ($\mu =+1.94\arcsec,
	  \sigma = 2.07\arcsec$).
	Middle graph: residuals versus visual magnitude of the asteroid;
	  the least-squares regression line is $R = -6.92 + (0.77\pm0.01)M_V$.
	Bottom graph: residuals versus time after magnitude correction
	  ($\mu = 0.00\arcsec, \sigma = 1.86\arcsec$).}
  \label{Fig1}
\end{figure}
 \item
  Each time bin of each group is inspected for outliers.
  The basic idea is that each bin consists of a population of "good" data
  having Gaussian distribution and a population of outliers.
  To remove these outliers, the kurtosis parameter~$k$ is used (see Carpino et
  al.~(\cite{Carp03}) for a definition).
  Its value is equal to 3 for a normal distribution, and substantially larger
  when outliers are present.
  The calculation of~$k$ and the outlier removal is done iteratively.
  The residual with the largest absolute deviation from the mean~$\mu$ is removed
  and $\mu$ and $k$ are recomputed.
  This is repeated until $k \le 3$ or, as a safety measure, until $k$ starts
  to increase rather than decrease.

\begin{table*}
\caption[]{Minor planets for which \emph{no} acceptable mass could be calculated.
  Those above the horizontal line gave negative masses and were removed,
  those below gave unacceptable results (see text) but were retained as perturbers
  in the force model with their estimated mass.
  The table gives: minor planet number and name, taxonomic class, adopted diameter
  and estimated mass.}
\label{NoMass}
\centering
\begin{tabular}{r@{ }l@{ }c@{ }c@{ }c@{ }|r@{ }l@{ }c@{ }c@{ }c@{ }@{ }|r@{ }l@{ }c@{ }c@{ }c}
 \hline \hline
  \multicolumn{2}{c}{Minor planet} & Tax. & Diam. & Est. mass &
  \multicolumn{2}{c}{Minor planet} & Tax. & Diam. & Est. mass &
  \multicolumn{2}{c}{Minor planet} & Tax. & Diam. & Est. mass \\
   & & class & (km) & $(10^{-10} M_{\sun})$ &
   & & class & (km) & $(10^{-10} M_{\sun})$ &
   & & class & (km) & $(10^{-10} M_{\sun})$\\
 \hline
    12 & Victoria	& S & 127 & 0.012 &  225 & Henrietta	& F & 109 & 0.007 &  602 & Marianna	& C & 127 & 0.010 \\
    30 & Urania		& S &~~98 & 0.006 &  233 & Asterope	& T & 100 & 0.005 &  675 & Ludmilla	& S & 109 & 0.008 \\
    36 & Atalante	& C & 103 & 0.005 &  247 & Eukrate	& C & 134 & 0.011 &  712 & Boliviana	& C & 120 & 0.008 \\
    47 & Aglaja		& C & 127 & 0.010 &  266 & Aline	& C & 109 & 0.006 &  713 & Luscinia	& C & 103 & 0.005 \\
    57 & Mnemosyne	& S & 123 & 0.011 &  276 & Adelheid	& C & 103 & 0.005 &  747 & Winchester	& P & 184 & 0.033 \\
    78 & Diana		& C & 121 & 0.008 &  303 & Josephina	& F & 106 & 0.006 &  748 & Simeisa	& P & 103 & 0.006 \\
    86 & Semele		& C & 115 & 0.007 &  361 & Bononia	& D & 142 & 0.015 &  769 & Tatjana	& X & 103 & 0.006 \\
    97 & Klotho		& M &~~83 & 0.006 &  386 & Siegena	& C & 169 & 0.023 &  776 & Berbericia	& C & 151 & 0.016 \\
    98 & Ianthe		& C & 111 & 0.007 &  420 & Bertholda	& P & 144 & 0.016 &  788 & Hohensteina	& C & 118 & 0.008 \\
   111 & Ate		& C & 129 & 0.010 &  455 & Bruchsalia	& C & 112 & 0.007 &  814 & Tauris	& C & 110 & 0.006 \\
   134 & Sophrosyne	& C & 112 & 0.007 &  471 & Papagena	& S & 134 & 0.014 &  895 & Helio	& F & 119 & 0.009 \\
   137 & Meliboea	& C & 144 & 0.014 &  489 & Comacina	& C & 139 & 0.013 &  909 & Ulla		& C & 116 & 0.008 \\
   140 & Siwa		& P & 110 & 0.007 &  498 & Tokia	& M &~~82 & 0.006 &  980 & Anacostia	& S &~~96 & 0.005 \\
   153 & Hilda		& P & 171 & 0.026 &  505 & Cava		& F & 105 & 0.006 & 1266 & Tone		& M &~~83 & 0.006 \\
   156 & Xanthippe	& C & 111 & 0.006 &  506 & Marion	& C & 106 & 0.006 & 1268 & Libya	& M &~~94 & 0.009 \\
   162 & Laurentia	& S & 104 & 0.007 &  508 & Princetonia	& C & 120 & 0.008 & 1390 & Abastumani	& C & 108 & 0.006 \\
   176 & Iduna		& G & 122 & 0.010 &  514 & Armida	& C & 106 & 0.006 & 1467 & Mashona	& C & 104 & 0.005 \\
   190 & Ismene		& P & 127 & 0.011 &  521 & Brixia	& C & 111 & 0.006 & 1754 & Cunningham	& M &~~80 & 0.006 \\
   201 & Penelope	& M &~~88 & 0.008 &  536 & Merapi	& C & 165 & 0.021 \\
   209 & Dido		& C & 124 & 0.009 &  545 & Messalina	& C & 113 & 0.007 \\
 \hline
    21 & Lutetia	& M & 105 & 0.013 &  181 & Eucharis	& S & 128 & 0.012 &  674 & Rachele	& S &~~96 & 0.005 \\	
    42 & Isis		& S & 100 & 0.006 &  227 & Philosophia	& F & 105 & 0.006 &  733 & Mocia	& C & 109 & 0.006 \\ 
    54 & Alexandra	& C & 142 & 0.014 &  229 & Adelinda	& B & 107 & 0.006 &  739 & Mandeville	& C & 104 & 0.005 \\
    63 & Ausonia	& S & 103 & 0.006 &  241 & Germania	& C & 192 & 0.034 &  751 & Fa\"ina	& C & 106 & 0.006 \\
    91 & Aegina		& C & 103 & 0.005 &  283 & Emma		& C & 135 & 0.012 &  762 & Pulcova	& F & 137 & 0.014 \\
    93 & Minerva	& C & 152 & 0.017 &  322 & Phaeo	& M &~~77 & 0.005 &  780 & Armenia	& F & 102 & 0.006 \\
   105 & Artemis	& C & 105 & 0.005 &  344 & Desiderata	& C & 126 & 0.009 &  849 & Ara		& M &~~84 & 0.007 \\ 
   135 & Hertha		& M &~~77 & 0.005 &  348 & May		& M &~~83 & 0.006 & 1093 & Freda	& C & 117 & 0.007 \\
   141 & Lumen		& C & 137 & 0.012 &  387 & Aquitania	& S & 101 & 0.006 & 1177 & Gonnessia	& M &~~92 & 0.009 \\
   146 & Lucina		& C & 132 & 0.011 &  410 & Chloris	& C & 119 & 0.008 & 1212 & Francette	& M &~~82 & 0.006 \\
   147 & Protogeneia	& C & 133 & 0.011 &  469 & Argentina	& X & 122 & 0.009 & 1269 & Rollandia	& D & 105 & 0.006 \\
   165 & Loreley	& C & 155 & 0.018 &  490 & Veritas	& C & 116 & 0.007 & 1911 & Schubart	& M &~~80 & 0.006 \\
   168 & Sibylla	& C & 144 & 0.014 &  570 & Kythera	& S & 110 & 0.008 \\
   175 & Andromache	& C & 115 & 0.007 &  595 & Polyxena	& C & 108 & 0.006 \\
\hline
\end{tabular}
\end{table*}

\item
  The complete set of residuals of each group (observatory code + observation
  technique) is then analysed for a (possible) magnitude effect.
  This is an observational bias whereby a measurement error is correlated
  with the object's brightness.
  Such an effect is to be expected for observations where the right ascension
  is measured by timing transits, such as meridian and most micrometer observations.
  It can also occur in photographic and CCD astrometry due to imperfections
  in the optical system, e.g. coma.
  To model the magnitude effect, a least-squares regression line of the form
	\[ R = a + bM_V \]
  is computed for both right ascension and declination residuals
  ($R$ is thus $\cos\delta \Delta\alpha$ or $\Delta\delta$).
  The quantity $M_V$ is the computed visual magnitude of the minor planet.
  In addition are computed: the total magnitude range $\Delta M$ of the set,
  the standard deviation $\sigma_b$ of $b$, the correlation coefficient
  $r$ (Pearson) and the statistic
	\[ t = r/\sigma_r = r \sqrt{(N-2)/(1-r^2)} \]
  where $N$ is the number of residuals.
  A magnitude effect is accepted as significant if:
	\[ \Delta M > 4.0 , \hspace{0.5em} |t| > 2 , \hspace{0.5em} \textup{and} \hspace{0.5em} |b|/\sigma_b  > 2 \]
  Fig. 1 gives an example of a substantial magnitude effect in right ascension.
 \item
  With outliers removed and after correcting for magnitude equation,
  the next step is then to compute for each time bin the values of the
  mean $\mu$ (the bias), the standard deviation $\sigma$ and the
  standard deviation of the mean
	\[ \sigma_\mu = \frac{\sigma}{\sqrt{N}} \]
  where $N$ is the number of remaining residuals in the bin.
  A bias is accepted as statistically significant when it exceeds 2 times
  its standard deviation ($|\mu|/\sigma_\mu > 2$).
  For the observations of code 248 Hipparcos the bias was set
  equal to zero, since they are by definition on the ICRS system.
\end{itemize}

The results are applied in \emph{P1} as follows.
Observations are considered as outliers, i.e. are given zero weight, if their
residuals deviate more than $3\sigma$ from the mean.
The weight of an observation is calculated as $w = 0.1/\sigma^2$ (unit weight thus
corresponds to $\sigma = \sqrt{0.1} \approx 0.32\arcsec$).
If there are $n$ observations of the same asteroid by the same observatory
(except for code 248~Hipparcos) on one night, then their weight is further
divided by $\sqrt{n}$, so that the total weight of the $n$ observations
is not $wn$ but $w\sqrt{n}$.
This procedure is a simple way to avoid overweighting them with respect to
older measurements. The reason is that their residuals are strongly correlated
because the observations were very probably made using the same reference stars.

\section{Results}
The computations with the compact storage formulation (Eq.~\ref{eq11},
with $N = 300\,000$ and $M = 105$) had indicated that the results were sensitive
to the weighting of the observations, and in particular to the recomputation
of the statistics.
Therefore, and in order to avoid a possible initial divergence of the
iterations, the first three iterations were done without changing the weights
or recomputing the perturber positions, and the recalculation of the observation
statistics ($P4$) was only set in after the 6th~iteration
(and then every 3 iterations).

After the main calculation was finished, a number of test computations
were performed to assess the sensitivity of the solution to the starting
values of the masses.

\subsection{Main calculation}
At the start of the computations the masses of the 230 perturbing asteroids
were set to their estimated values.
After five iterations it turned out that 58 of them persistently yielded
negative mass values in the first solution of the (unmodified) normal equations;
they are listed in Table~\ref{NoMass} (upper part).
It was decided to remove them and to restart the computations with the 172
remaining perturbing asteroids, partly to reduce computing time
	\footnote{For the calculations with 230 asteroid masses, the total reported
	CPU time of the programme \emph{P1} was about 96.5~hours per iteration,
	slightly over 59~hours with 172 perturbers.}.

Due to the improvement of the $N$ orbits and the reweighting of the observations
at each iteration, the mean error of one observation decreased to
$0.23\arcsec$ and the total number of equations of condition used in the
solution increased from 164.28 to 168.51 million.
Convergence for most masses was reached after 14 iterations, i.e. after that
their values started to oscillate by a few units of the 4th~decimal.
The mass of Pallas however showed a general decline and later started
fluctuating by one or two units of the \emph{3rd}~decimal.
The calculations were finally stopped after the 29th iteration; the total number
of linear equations used in the last iteration was 168\,506\,583.

\begin{table}
 \caption{Twenty closest approaches between test and perturbing asteroids.}
 \label{MinDist}
 \centering
\begin{tabular}{r@{ }lr@{ }lc}
 \hline \hline
 \multicolumn{2}{c}{Perturbing asteroid} & \multicolumn{2}{c}{Test asteroid} & Min. dist \\
   & & & & (au)\\
 \hline
  324 & Bamberga   & 310585 & 2001 UU$_{20}$  & 0.00006 \\
   50 & Virginia   &  79009 & 4707 P-L	      & 0.00014 \\
   20 & Massalia   & 336058 & 2008 EU$_{36}$  & 0.00015 \\
  110 & Lydia      &  66778 & 1999 TL$_{221}$ & 0.00016 \\
  324 & Bamberga   & 121416 & 1999 TG$_{146}$ & 0.00019 \\
   96 & Aegle      & 194481 & 2001 WA$_{37}$  & 0.00019 \\
   49 & Pales      & 145746 & 1995 UR$_{18}$  & 0.00023 \\
  106 & Dione      & 233456 & 2006 JS$_{38}$  & 0.00026 \\
  426 & Hippos     & 308965 & 2006 TG$_{67}$  & 0.00033 \\
  203 & Pompeja    & 204819 & 2007 PU$_{1}$   & 0.00034 \\
   40 & Harmonia   & 272519 & 2005 UA$_{256}$ & 0.00035 \\
  187 & Lamberta   & 206779 & 2004 CR$_{97}$  & 0.00035 \\
   56 & Melete     &  22960 & 1999 UE$_{4}$   & 0.00035 \\
   38 & Leda       &  47863 & 2000 EC$_{180}$ & 0.00035 \\
   15 & Eunomia    &  50278 & 2000 CZ$_{12}$  & 0.00038 \\
  128 & Nemesis    & 272917 & 2006 BR$_{180}$ & 0.00039 \\
  211 & Isolda     & 257257 & 2009 FB$_{43}$  & 0.00042 \\
   11 & Parthenope & 334240 & 2001 TW$_{96}$  & 0.00045 \\
  110 & Lydia      &  70797 & 1999 VF$_{54}$  & 0.00046 \\
   74 & Galatea    &  63721 & 2001 QH$_{234}$ & 0.00047 \\
 \hline
 \end{tabular}
\end{table}

For 40 of the 172 perturbing asteroids no acceptable mass could be derived
according to the significance and density criteria given above.
They are listed in Table~\ref{NoMass} (lower part) together with their
taxonomic class, estimated diameter and mass used in the dynamical model.
Table~\ref{MinDist} gives the 20~closest approaches between test and
perturbing asteroids.
Table~\ref{Masses} gives the 132 asteroids with acceptable astrometric masses,
together with their adopted (equivalent) diameter, taxonomic class
and bulk density derived from these quantities.
The results for 50 of them appear to be new; they are indicated by an
asterisk in the table.

\begin{table}
 \caption{Ten largest correlations between the masses of perturbers and the semi-major
	axes of the test asteroids.}
 \label{Correl1}
 \centering
\begin{tabular}{r@{ }lr@{ }lc}
 \hline \hline
 \multicolumn{2}{c}{Perturbing asteroid} & \multicolumn{2}{c}{Test asteroid} & Corr. coeff.\\
 \hline
    2 & Pallas    &  58243 & 1993 NG$_{1}$   & $+0.99$ \\
    2 & Pallas    & 133978 & 2004 TK$_{242}$ & $+0.97$ \\
   15 & Eunomia	  &  50278 & 2000 CZ$_{12}$  & $-0.96$ \\
   49 & Pales     & 145746 & 1995 UR$_{18}$  & $+0.96$ \\
   88 & Thisbe    &  70336 & 1999 RO$_{169}$ & $+0.96$ \\
  196 & Philomela & 104441 & 2000 GQ         & $-0.95$ \\
  419 & Aurelia   & 118346 & 1999 CA$_{49}$  & $-0.95$ \\
    9 & Metis     &  36653 & 2000 QF$_{200}$ & $-0.94$ \\
  106 & Dione     & 233456 & 2006 JS$_{38}$  & $-0.93$ \\
  423 & Diotima   &  72812 & 2001 GB$_{8}$   & $-0.93$ \\
 \hline
 \end{tabular}
\end{table}

\begin{table}
 \caption{Ten largest mutual correlations between the perturber masses.}
 \label{Correl2}
 \centering
 \begin{tabular}{r@{ }lr@{ }lc}
 \hline \hline
 \multicolumn{2}{c}{Perturber 1} & \multicolumn{2}{c}{Perturber 2} & Corr. coeff. \\
 \hline
   17 & Thetis     & 145 & Adeona     & $+0.56$ \\
    4 & Vesta      &  11 & Parthenope & $-0.45$ \\
   17 & Thetis     & 275 & Sapientia  & $+0.39$ \\
   11 & Parthenope &  40 & Harmonia   & $-0.24$ \\
   88 & Thisbe     &  89 & Julia      & $+0.22$ \\
   49 & Pales      & 175 & Andromache & $-0.18$ \\
    5 & Astraea    & 356 & Liguria    & $-0.17$ \\
    2 & Pallas     & 203 & Pompeja    & $-0.17$ \\
   76 & Freia      & 203 & Pompeja    & $-0.16$ \\
  139 & Juewa      & 212 & Medea      & $+0.16$ \\
 \hline
 \end{tabular}
\end{table}

As is to be expected, there are high correlations between the perturber masses
and the semi-major axes of some of the test asteroids.
The 10 largest values are given in Table~\ref{Correl1}; all of them involve
high-numbered test asteroids.
The mutual correlations between the $M$ masses turn out to be generally low; the
10 largest values are listed in Table~\ref{Correl2}.
All correlations between the 10 \emph{largest} asteroid masses are $\le 0.05$.

\subsection{Influence of initial assumed masses}
After concluding the main calculation, the perturber masses were set to zero
and the iterations were resumed \emph{without} reweighting the observations,
i.e. with the iteration sequence $P1+P2+P3$.
After three iterations, the masses were back to the results of the main
calculation to within four decimal places.
Next, the masses were set to twice their assumed values and the calculations
were continued. Again, three iterations sufficed.
This indicates that the least-squares algorithm is rather insensitive to
the starting mass values.

Then, the \emph{main} calculations as descibed above were repeated twice,
first with all perturber masses initially set to zero and then set
to twice their estimated values, but this time with reweighting applied.
The statistical analysis obtained at the end of the previous set of iterations
was used at the beginning of the new test series.
Convergence was deemed to be reached in both cases after 20 iterations; the number
of linear equations in the last iteration steps was less than in the main
calculation: $-20\,222$ ($-0.012\%$) and $-30\,237$ ($-0.018\%$) respectively.
In Table~\ref{Masses} the mass, standard deviation and significance are
from the main iteration.
The mass differences between the last two test iterations and the main one are
also included (column~$\Delta m_0$ for the zero and column~$\Delta m_2$ for the
double starting mass values).

\section{Discussion}
The masses given in Table~\ref{Masses} are presented ''as such'', i.e. as the
outcome of the least-squares adjustment described above.
The bulk density is a quantity derived from the mass and the assumed diameter
and its uncertainty is thus affected by the uncertainties of both the
mass and the diameter.
No physical interpretation of the densities will be attempted here.
Some results conform to what is to be expected for their taxonomic class,
while others are higher or lower.
One should bear in mind that the taxonomic class is derived from the spectral
properties of the surface of an asteroid and as such does not say much about
its internal composition or porosity.

The mass and size of 4~Vesta is accurately known from the orbit of the
Dawn spacecraft. Russell et al.~(\cite{Russ2012}) give a value of
$2.59076 \pm 0.00001 \times 10^{20}$~kg corresponding to
$1.30260 \pm 0.00001 \times 10^{-10} M_{\sun}$, an equivalent diameter
of 523.15~km and a density of $3.456~\mathrm{g/cm^3}$.
The result from this study, to 4 decimal places, is $1.2995 \pm 0.0015$
yielding a density of $3.448~\mathrm{g/cm^3}$, in good agreement with
the Dawn result.

It would be a constructive achievement to develop an objective criterion to
exclude test asteroids that do not significantly contribute to the solution.
The correlations between the masses and the semi-major axes come to mind as
an obvious choice.
Using the sum of the absolute values of the $M$~correlation coefficients
as an indicator for the usefulness of a test asteroid, test calculations
were made with $N = 10,000$ and $M = 10$ (the 10 largest masses).
With a limit for the sum of 0.01, about 15\% were excluded and the average
difference in the 10~masses was 0.0003; this became 48\% and 0.0012
with 0.02 as the limit.
It seems worthwhile to test this criterion on the full set of test asteroids.

In general, the two test calculations clearly show that the most influential
factor is the treatment of the observations (weighting and outlier rejection).
In that respect, it should be remarked that the observations reduced with the
USNO~A1.0, A~2.0 and B~1.0 star catalogues have not been corrected for
the systematic biases as described in Chesley et al.~(\cite{Ches10})
or in R\"oser et al.~(\cite{Rose10}).
It should be worthwile to include these in future work.

The planetary ephemeris uncertainties are negligible in this case,
certainly the positional errors. The planetary masses come from spacecraft
encounters, and DE421/422 is up-to-date.
The best asteroid-produced planetary mass of Jupiter, for example, was
inaccurate by about 6 parts in $10^6$; the spacecraft-produced values are
two orders of magnitude more accurate.

The Gaia satellite, launched at the end of 2013, will in the course of its
5 year nominal duration systematically observe $\sim$300\,000 asteroids
brighter than $V = 20$ with an accuracy of 0.3--5 mas.
This will enable the determination of the masses of about 150 individual
asteroids from close encounters, $\sim$100 with an accuracy of $\le 50\%$
and $\sim$50 with an accuracy of $\le 10\%$ (Mouret et al.~\cite{Mour2008}).
The Gaia star catalogue will enable the reduction of future, and the
re-reduction of old observations with a very high level of accuracy
(about 10~mas, see Desmars et al.~\cite{Desm2013}),
thus considerably increasing the accuracy of simultaneous asteroid
mass determinations.

Finally, Table~\ref{Overview} gives an overview of mass results from recent
determinations for those minor planets given in Table~\ref{Masses}, and
published in roughly the last ten years.
For a more complete list inlcuding older determinations, the reader is
referred to Zielenbach~(\cite{Ziel2011}).
Besides dynamical determinations using asteroids as test objects, there are
also determinations from natural satellites and from planetary ephemerides.
Note that these use a completely different type of observations.

\begin{acknowledgements}
I am most indebted to E. Myles Standish for his keen interest in the subject,
his willingness to critically review this paper and for bringing in his
expertise where my mathematical knowledge failed.
\end{acknowledgements}

\begin{longtab}
\begin{longtable}{r@{ }lccrrrrcc}
 \caption{\label{Masses} Masses of 132 asteroids obtained from the simultaneous
  solution. The table gives: minor planet number and name ($^*$ = new mass
  determination), mass, standard deviation and significance, mass difference
  when starting from masses equal to zero~($\Delta m_0$) and from masses
  twice~($\Delta m_2$) the assumed values, adopted equivalent diameter,
  taxonomic class and resulting bulk density. Standard deviations printed as
  ''0.000'' indicate values $< 0.0005$; their exact value can be inferred from
  the values of the mass and its significance.} \\
 \hline \hline
   \multicolumn{2}{c}{Minor planet} & Mass & St. dev. & Signi- &$\Delta m_0$ & $\Delta m_2$ &
		Diam. & Taxon. & Dens. \\
   & & \multicolumn{2}{c}{$(10^{-10} M_{\sun})$} & ficance & \multicolumn{2}{c}{$(10^{-13} M_{\sun})$}
		& (km) & class & $(\mathrm{g/cm^3})$ \\
 \hline
 \endfirsthead
 \caption{continued.}\\
 \hline \hline
   \multicolumn{2}{c}{Minor planet} & Mass & St. dev. & Signi- &$\Delta m_0$ & $\Delta m_2$ &
		Diam. & Taxon. & Dens. \\
   & & \multicolumn{2}{c}{$(10^{-10} M_{\sun})$} & ficance & \multicolumn{2}{c}{$(10^{-13} M_{\sun})$}
		& (km) & class & $(\mathrm{g/cm^3})$ \\
 \hline
 \endhead
 \hline
 \endfoot
     1 & Ceres            & 4.748 & 0.003 &1561.4 & $+1.1$ & $+0.5$ & 953 & G & 2.09 \\
     2 & Pallas           & 1.007 & 0.011 &  88.1 & $-2.9$ & $-2.7$ & 531 & B & 2.56 \\
     3 & Juno             & 0.126 & 0.004 &  31.6 & $-1.5$ & $-0.5$ & 258 & S & 2.81 \\
     4 & Vesta            & 1.300 & 0.002 & 847.2 & $-0.2$ & $-0.1$ & 523 & V & 3.45 \\
     5 & Astraea          & 0.033 & 0.003 &  12.4 & $+0.3$ & $+0.2$ & 127 & S & 6.10 \\
     6 & Hebe             & 0.045 & 0.003 &  13.4 & $-1.2$ & $-0.7$ & 186 & S & 2.64 \\
     7 & Iris             & 0.070 & 0.002 &  34.9 & $-0.5$ & $-0.3$ & 213 & S & 2.78 \\
     8 & Flora            & 0.029 & 0.002 &  17.0 & $-2.1$ & $-2.5$ & 161 & S & 2.65 \\
     9 & Metis            & 0.025 & 0.002 &  15.9 & $+0.4$ & $+0.3$ & 174 & S & 1.84 \\
    10 & Hygiea           & 0.410 & 0.004 & 115.8 & $-0.3$ & $-0.3$ & 431 & C & 1.95 \\
    11 & Parthenope       & 0.025 & 0.000 &  69.5 & $ 0.0$ & $+0.1$ & 153 & S & 2.67 \\
    13 & Egeria           & 0.047 & 0.004 &  12.0 & $+0.2$ & $+0.1$ & 208 & G & 1.97 \\
    14 & Irene            & 0.028 & 0.003 &   9.6 & $-1.1$ & $-1.1$ & 152 & S & 2.97 \\
    15 & Eunomia          & 0.146 & 0.002 &  64.3 & $-0.4$ & $-0.3$ & 268 & S & 2.88 \\
    16 & Psyche           & 0.117 & 0.002 &  56.6 & $-0.7$ & $-0.5$ & 186 & M & 6.89 \\
    17 & Thetis           & 0.004 & 0.000 &  12.5 & $-0.1$ & $-0.1$ &  98 & S & 1.51 \\
    18 & Melpomene        & 0.027 & 0.003 &  10.2 & $+0.9$ & $+0.6$ & 150 & S & 3.05 \\
    19 & Fortuna          & 0.045 & 0.001 &  35.3 & $-0.3$ & $-1.4$ & 208 & G & 1.89 \\
    20 & Massalia         & 0.024 & 0.001 &  19.7 & $-0.3$ & $ 0.0$ & 145 & S & 3.01 \\
    22 & Kalliope         & 0.024 & 0.004 &   6.4 & $+0.3$ & $-0.1$ & 175 & M & 1.73 \\
    23 & Thalia           & 0.023 & 0.003 &   7.5 & $+0.8$ & $+0.9$ & 108 & S & 7.06 \\
    24 & Themis           & 0.051 & 0.003 &  16.3 & $-0.7$ & $-0.3$ & 198 & C & 2.50 \\
    27 & Euterpe          & 0.010 & 0.002 &   5.0 & $+0.2$ & $+0.3$ & 118 & S & 2.23 \\
    28 & Bellona          & 0.008 & 0.003 &   2.9 & $-0.5$ & $ 0.0$ & 121 & S & 1.66 \\
    29 & Amphitrite       & 0.067 & 0.002 &  44.0 & $+0.2$ & $+0.2$ & 212 & S & 2.67 \\
    31 & Euphrosyne       & 0.087 & 0.007 &  12.7 & $-0.1$ & $-0.9$ & 256 & C & 1.96 \\
    34 & Circe$^*$        & 0.021 & 0.002 &   9.0 & $-0.7$ & $-0.8$ & 113 & C & 5.40 \\
    37 & Fides$^*$        & 0.012 & 0.002 &   6.5 & $-0.5$ & $-0.4$ & 108 & S & 3.49 \\
    38 & Leda$^*$         & 0.016 & 0.003 &   6.2 & $+1.1$ & $+1.4$ & 116 & C & 3.79 \\
    39 & Laetitia         & 0.022 & 0.003 &   7.5 & $-0.4$ & $-0.1$ & 163 & S & 1.93 \\
    40 & Harmonia$^*$     & 0.020 & 0.002 &  10.5 & $+0.2$ & $+0.5$ & 120 & S & 4.35 \\
    41 & Daphne           & 0.047 & 0.007 &   7.2 & $-0.9$ & $-0.6$ & 177 & C & 3.25 \\
    45 & Eugenia          & 0.032 & 0.002 &  17.0 & $-1.2$ & $-1.3$ & 193 & F & 1.69 \\
    46 & Hestia           & 0.005 & 0.002 &   2.8 & $ 0.0$ & $ 0.0$ & 124 & P & 0.95 \\
    48 & Doris            & 0.039 & 0.003 &  11.8 & $-0.6$ & $-0.6$ & 222 & C & 1.35 \\
    49 & Pales            & 0.027 & 0.003 &  10.5 & $-0.4$ & $-0.7$ & 150 & C & 3.09 \\
    50 & Virginia$^*$     & 0.003 & 0.000 &   6.5 & $ 0.0$ & $+0.1$ & 100 & X & 1.02 \\
    51 & Nemausa          & 0.014 & 0.002 &   8.1 & $-0.1$ & $-0.5$ & 145 & C & 1.81 \\
    52 & Europa           & 0.135 & 0.003 &  43.5 & $-0.5$ & $-0.7$ & 315 & C & 1.82 \\
    53 & Kalypso$^*$      & 0.008 & 0.002 &   3.7 & $ 0.0$ & $+0.2$ & 115 & X & 2.12 \\
    56 & Melete$^*$       & 0.019 & 0.003 &   7.1 & $-0.9$ & $-1.0$ & 129 & P & 3.33 \\
    59 & Elpis            & 0.017 & 0.003 &   6.4 & $-0.4$ & $-0.5$ & 166 & C & 1.43 \\
    65 & Cybele           & 0.089 & 0.004 &  24.0 & $-0.4$ & $-0.9$ & 273 & P & 1.66 \\
    68 & Leto             & 0.021 & 0.003 &   7.9 & $+0.9$ & $+0.6$ & 129 & S & 3.72 \\
    69 & Hesperia         & 0.047 & 0.004 &  11.1 & $-0.4$ & $-1.0$ & 138 & M & 6.82 \\
    70 & Panopaea$^*$     & 0.017 & 0.002 &   8.3 & $+0.1$ & $+0.1$ & 139 & C & 2.38 \\
    74 & Galatea$^*$      & 0.005 & 0.002 &   2.3 & $+0.4$ & $+0.3$ & 119 & C & 1.24 \\
    76 & Freia$^*$        & 0.038 & 0.004 &   9.0 & $+0.4$ & $+0.1$ & 159 & P & 3.64 \\
    81 & Terpsichore$^*$  & 0.016 & 0.001 &  11.5 & $ 0.0$ & $-0.1$ & 122 & C & 3.31 \\
    85 & Io               & 0.012 & 0.003 &   3.5 & $-0.3$ & $ 0.0$ & 164 & F & 1.02 \\
    87 & Sylvia           & 0.108 & 0.010 &  11.0 & $-0.3$ & $-3.2$ & 286 & P & 1.76 \\
    88 & Thisbe           & 0.051 & 0.002 &  20.5 & $+0.4$ & $+0.4$ & 225 & C & 1.69 \\
    89 & Julia            & 0.043 & 0.003 &  13.1 & $-0.8$ & $-1.0$ & 148 & S & 5.05 \\
    92 & Undina$^*$       & 0.037 & 0.005 &   7.0 & $-0.7$ & $+0.2$ & 126 & X & 6.93 \\
    94 & Aurora           & 0.009 & 0.003 &   3.2 & $+0.1$ & $ 0.0$ & 178 & C & 0.59 \\
    95 & Arethusa$^*$     & 0.021 & 0.004 &   5.6 & $+0.3$ & $+0.2$ & 145 & C & 2.61 \\
    96 & Aegle            & 0.057 & 0.003 &  17.9 & $+0.2$ & $+0.6$ & 170 & T & 4.38 \\
   104 & Klymene$^*$      & 0.009 & 0.002 &   3.7 & $ 0.0$ & $-0.1$ & 126 & C & 1.66 \\
   106 & Dione            & 0.018 & 0.003 &   5.5 & $+0.3$ & $+0.3$ & 136 & G & 2.77 \\
   107 & Camilla          & 0.056 & 0.009 &   6.4 & $-0.2$ & $-1.4$ & 259 & C & 1.22 \\
   110 & Lydia$^*$        & 0.007 & 0.000 &  22.4 & $ 0.0$ & $ 0.0$ &  89 & M & 3.89 \\
   114 & Kassandra$^*$    & 0.005 & 0.002 &   2.1 & $+0.4$ & $-0.1$ & 100 & T & 1.81 \\
   117 & Lomia            & 0.017 & 0.004 &   4.0 & $-0.7$ & $-0.9$ & 173 & C & 1.24 \\
   120 & Lachesis$^*$     & 0.032 & 0.004 &   7.7 & $-0.9$ & $-0.5$ & 163 & C & 2.83 \\
   121 & Hermione         & 0.016 & 0.002 &   7.2 & $-0.4$ & $-0.5$ & 187 & C & 0.96 \\
   128 & Nemesis          & 0.042 & 0.003 &  16.3 & $-0.3$ & $-0.2$ & 188 & C & 2.37 \\
   129 & Antigone         & 0.032 & 0.004 &   9.1 & $-0.2$ & $-0.3$ & 130 & M & 5.67 \\
   130 & Elektra          & 0.070 & 0.008 &   8.9 & $-0.7$ & $-1.5$ & 215 & G & 2.67 \\
   139 & Juewa            & 0.030 & 0.003 &   9.0 & $+0.4$ & $+0.3$ & 164 & C & 2.57 \\
   144 & Vibilia          & 0.017 & 0.002 &   7.6 & $ 0.0$ & $+0.1$ & 152 & C & 1.83 \\
   145 & Adeona           & 0.014 & 0.003 &   4.6 & $+0.6$ & $+0.9$ & 151 & C & 1.55 \\
   150 & Nuwa             & 0.017 & 0.003 &   6.3 & $-0.3$ & $-0.5$ & 137 & C & 2.51 \\
   154 & Bertha           & 0.081 & 0.008 &   9.8 & $+1.6$ & $+0.8$ & 189 & C & 4.57 \\
   159 & Aemilia$^*$      & 0.021 & 0.003 &   7.4 & $-0.7$ & $-0.5$ & 127 & C & 3.94 \\
   164 & Eva$^*$          & 0.011 & 0.006 &   1.6 & $-0.6$ & $-0.1$ & 109 & C & 3.10 \\
   171 & Ophelia$^*$      & 0.006 & 0.003 &   2.2 & $+0.2$ & $+0.1$ & 104 & C & 2.13 \\
   173 & Ino              & 0.016 & 0.003 &   5.8 & $-0.5$ & $-2.2$ & 154 & C & 1.68 \\
   185 & Eunike$^*$       & 0.021 & 0.006 &   3.2 & $+0.9$ & $+1.1$ & 156 & C & 2.08 \\
   187 & Lamberta         & 0.011 & 0.004 &   3.1 & $+0.6$ & $+0.2$ & 133 & C & 1.72 \\
   194 & Prokne           & 0.023 & 0.003 &   6.9 & $+0.3$ & $+0.4$ & 169 & C & 1.83 \\
   196 & Philomela        & 0.027 & 0.004 &   6.3 & $ 0.0$ & $-0.5$ & 158 & S & 2.64 \\
   200 & Dynamene$^*$     & 0.020 & 0.003 &   7.9 & $-0.6$ & $-0.8$ & 131 & C & 3.42 \\
   202 & Chryse\"is$^*$   & 0.015 & 0.002 &   7.4 & $-0.2$ & $-0.2$ &  98 & S & 5.98 \\
   203 & Pompeja$^*$      & 0.006 & 0.002 &   4.1 & $-0.1$ & $-0.2$ & 111 & D & 1.75 \\
   206 & Hersilia$^*$     & 0.018 & 0.002 &  11.9 & $-0.4$ & $-0.4$ & 105 & C & 6.05 \\
   211 & Isolda           & 0.015 & 0.003 &   6.0 & $+0.1$ & $+0.3$ & 143 & C & 1.94 \\
   212 & Medea            & 0.011 & 0.001 &  10.2 & $+0.1$ & $+0.1$ & 136 & D & 1.65 \\
   216 & Kleopatra        & 0.026 & 0.004 &   6.4 & $+0.9$ & $+0.3$ & 138 & M & 3.83 \\
   221 & Eos$^*$          & 0.012 & 0.003 &   3.7 & $+0.6$ & $+0.5$ &  96 & S & 5.40 \\
   230 & Athamantis$^*$   & 0.012 & 0.002 &   5.0 & $+0.2$ & $-0.4$ & 109 & S & 3.57 \\
   238 & Hypatia          & 0.015 & 0.003 &   4.6 & $-0.1$ & $ 0.0$ & 145 & C & 1.83 \\
   250 & Bettina$^*$      & 0.013 & 0.005 &   2.4 & $-0.2$ & $+0.2$ & 121 & M & 2.69 \\
   259 & Aletheia         & 0.065 & 0.005 &  13.8 & $-0.4$ & $-1.0$ & 207 & C & 2.79 \\
   268 & Adorea           & 0.016 & 0.002 &   6.8 & $+0.1$ & $+0.1$ & 141 & F & 2.14 \\
   275 & Sapientia$^*$    & 0.013 & 0.003 &   5.0 & $+0.5$ & $-0.1$ & 103 & C & 4.50 \\
   308 & Polyxo$^*$       & 0.024 & 0.002 &  14.5 & $-0.8$ & $-0.7$ & 144 & T & 2.99 \\
   324 & Bamberga         & 0.044 & 0.002 &  18.7 & $-0.8$ & $-0.8$ & 212 & C & 1.77 \\
   328 & Gudrun           & 0.029 & 0.007 &   4.3 & $-1.4$ & $-1.3$ & 116 & S & 6.94 \\
   334 & Chicago$^*$      & 0.094 & 0.018 &   5.1 & $+1.2$ & $-0.8$ & 173 & C & 6.92 \\
   349 & Dembowska        & 0.014 & 0.003 &   4.8 & $+0.3$ & $ 0.0$ & 143 & R & 1.82 \\
   354 & Eleonora         & 0.089 & 0.006 &  15.4 & $+0.2$ & $+1.0$ & 165 & S & 7.53 \\
   356 & Liguria$^*$      & 0.014 & 0.002 &   8.5 & $+0.4$ & $+0.4$ & 131 & C & 2.43 \\
   360 & Carlova$^*$      & 0.034 & 0.003 &  10.6 & $ 0.0$ & $-1.1$ & 133 & C & 5.60 \\
   372 & Palma$^*$        & 0.032 & 0.005 &   6.1 & $+0.3$ & $+0.3$ & 191 & B & 1.74 \\
   375 & Ursula           & 0.059 & 0.006 &  10.2 & $-0.1$ & $-1.0$ & 204 & C & 2.65 \\
   381 & Myrrha$^*$       & 0.035 & 0.005 &   7.0 & $+0.2$ & $ 0.0$ & 129 & C & 6.18 \\
   388 & Charybdis$^*$    & 0.027 & 0.003 &   8.0 & $-0.2$ & $-0.4$ & 124 & C & 5.30 \\
   405 & Thia$^*$         & 0.011 & 0.005 &   2.0 & $+0.8$ & $+1.2$ & 125 & C & 2.09 \\
   409 & Aspasia          & 0.027 & 0.003 &   9.5 & $+0.9$ & $+0.9$ & 163 & C & 2.41 \\
   419 & Aurelia$^*$      & 0.022 & 0.002 &  11.6 & $-0.3$ & $ 0.0$ & 129 & F & 3.87 \\
   423 & Diotima          & 0.048 & 0.004 &  11.6 & $-0.1$ & $+0.3$ & 170 & C & 3.73 \\
   426 & Hippo$^*$        & 0.011 & 0.003 &   3.9 & $+0.4$ & $-0.1$ & 116 & F & 2.56 \\
   444 & Gyptis$^*$       & 0.007 & 0.003 &   2.3 & $+1.7$ & $+1.8$ & 163 & C & 0.62 \\
   445 & Edna$^*$         & 0.021 & 0.006 &   3.3 & $-0.9$ & $-0.8$ & 106 & C & 6.70 \\
   451 & Patientia        & 0.068 & 0.005 &  14.6 & $+1.2$ & $+1.3$ & 225 & C & 2.27 \\
   476 & Hedwig$^*$       & 0.007 & 0.002 &   3.0 & $ 0.0$ & $+0.1$ & 117 & P & 1.65 \\
   488 & Kreusa           & 0.037 & 0.005 &   7.9 & $-0.4$ & $-0.4$ & 150 & C & 4.14 \\
   511 & Davida           & 0.151 & 0.005 &  30.9 & $+0.0$ & $+0.0$ & 289 & C & 2.37 \\
   532 & Herculina        & 0.104 & 0.004 &  26.5 & $+0.8$ & $+0.7$ & 222 & S & 3.60 \\
   554 & Peraga$^*$       & 0.004 & 0.001 &   3.0 & $-0.1$ & $-0.1$ & 105 & F & 1.30 \\
   566 & Stereoskopia$^*$ & 0.031 & 0.005 &   6.1 & $-1.0$ & $-1.0$ & 134 & C & 4.97 \\
   596 & Scheila$^*$      & 0.027 & 0.006 &   4.5 & $-0.6$ & $-0.3$ & 114 & P & 6.87 \\
   618 & Elfriede$^*$     & 0.016 & 0.005 &   3.1 & $-0.2$ & $-0.1$ & 131 & C & 2.62 \\
   654 & Zelinda$^*$      & 0.034 & 0.005 &   6.4 & $+0.2$ & $+0.1$ & 120 & C & 7.38 \\
   690 & Wratislavia      & 0.011 & 0.005 &   2.2 & $-0.6$ & $ 0.0$ & 135 & C & 1.76 \\
   702 & Alauda           & 0.066 & 0.010 &   6.7 & $+0.0$ & $ 0.0$ & 207 & C & 2.82 \\
   704 & Interamnia       & 0.137 & 0.006 &  24.0 & $-0.8$ & $ 0.0$ & 319 & F & 1.60 \\
   705 & Erminia$^*$      & 0.039 & 0.006 &   6.4 & $-3.6$ & $-2.0$ & 131 & C & 6.63 \\
   772 & Tanete$^*$       & 0.058 & 0.009 &   6.8 & $+0.5$ & $-0.4$ & 143 & C & 7.63 \\
   790 & Pretoria         & 0.049 & 0.009 &   5.1 & $+2.7$ & $+3.2$ & 170 & P & 3.74 \\
   804 & Hispania         & 0.014 & 0.003 &   4.9 & $ 0.0$ & $-0.3$ & 147 & P & 1.69 \\
  1428 & Mombasa$^*$      & 0.019 & 0.004 &   4.4 & $-0.3$ & $-0.5$ & 127 & S & 3.54 \\
\end{longtable}
\end{longtab}

\onecolumn
\onltab{
\begin{longtab}
\begin{longtable}{r@{ }lllr}
 \caption{\label{Overview} Mass determinations of minor planets from recent
   dynamical results and latest planetary ephemerides.
   The table gives: minor planet number and name, mass and standard deviation
   and reference ($^p$ = planetary ephemerides, $^s$ = binary or natural satllites).} \\
 \hline \hline
 \multicolumn{2}{c}{Minor planet} & Mass & St. Dev. & Ref. \\
  & & \multicolumn{2}{c}{($10^{-10} M_{\sun})$} \\
 \hline
 \endfirsthead
 \caption{continued.} \\
 \hline \hline
 \multicolumn{2}{c}{Minor planet} & Mass & St. Dev. & Ref. \\
  & & \multicolumn{2}{c}{($10^{-10} M_{\sun})$} \\
 \hline
 \endhead
 \hline
 \endfoot
   1 & Ceres           & 4.759 & 0.023 & $ 1~~$ \\
     &                 & 4.70  & 0.04  & $ 2~~$ \\
     &                 & 4.736 & 0.026 & $ 3~~$ \\
     &                 & 4.75  & 0.03  & $ 4~~$ \\
     &                 & 4.685 &       & $ 5^p$ \\
     &                 & 4.712 & 0.006 & $ 6~~$ \\
     &                 & 4.757 & 0.007 & $ 8~~$ \\
     &                 & 4.679 & 0.033 & $ 9~~$ \\
     &                 & 4.735 & 0.013 & $10~~$ \\
     &                 & 4.731 & 0.029 & $37^p$ \\
     &                 & 4.673 & 0.019 & $ 7^p$ \\
     &                 & 4.748 & 0.003 & $99~~$ \\
   2 & Pallas          & 1.21  & 0.26  & $ 2~~$ \\
     &                 & 1.17  & 0.03  & $11~~$ \\
     &                 & 1.06  & 0.13  & $ 4~~$ \\
     &                 & 1.010 &       & $ 5^p$ \\
     &                 & 1.027 & 0.007 & $ 6~~$ \\
     &                 & 1.010 & 0.065 & $ 8~~$ \\
     &                 & 1.034 & 0.069 & $10~~$ \\
     &                 & 1.017 & 0.021 & $37^p$ \\
     &                 & 1.127 & 0.016 & $ 7^p$ \\
     &                 & 1.007 & 0.011 & $99~~$ \\
   3 & Juno            & 0.209 & 0.035 & $14~~$ \\
     &                 & 0.144 & 0.023 & $ 8~~$ \\
     &                 & 0.156 & 0.016 & $10~~$ \\
     &                 & 0.135 & 0.008 & $37^p$ \\
     &                 & 0.118 & 0.006 & $ 7^p$ \\
     &                 & 0.126 & 0.004 & $99~~$ \\
   4 & Vesta           & 1.36  & 0.05  & $ 2~~$ \\
     &                 & 1.306 & 0.016 & $15~~$ \\
     &                 & 1.282 & 0.018 & $16~~$ \\
     &                 & 1.34  & 0.01  & $ 4~~$ \\
     &                 & 1.328 &       & $ 5^p$ \\
     &                 & 1.344 & 0.003 & $ 6~~$ \\
     &                 &  1.30 &  0.00 & $ 2~~$ \\
     &                 & 1.300 & 0.005 & $ 8~~$ \\
     &                 & 1.303 & 0.007 & $10~~$ \\
     &                 & 1.303 & 0.000 & $34^s$ \\
     &                 & 1.305 & 0.008 & $37^p$ \\
     &                 & 1.303 & 0.009 & $ 7^p$ \\
     &                 & 1.300 & 0.002 & $99~~$ \\
   5 & Astraea         & 0.043 & 0.011 & $10~~$ \\
     &                 & 0.033 & 0.003 & $99~~$ \\
   6 & Hebe            & 0.069 & 0.022 & $17~~$ \\
     &                 & 0.069 & 0.009 & $14~~$ \\
     &                 & 0.076 & 0.014 & $ 4~~$ \\
     &                 & 0.064 & 0.007 & $ 8~~$ \\
     &                 & 0.077 & 0.012 & $10~~$ \\
     &                 & 0.042 & 0.010 & $37^p$ \\
     &                 & 0.071 & 0.008 & $ 7^p$ \\
     &                 & 0.045 & 0.003 & $99~~$ \\
   7 & Iris            & 0.141 & 0.014 & $14~~$ \\
     &                 & 0.24  & 0.03  & $19~~$ \\
     &                 & 0.068 & 0.019 & $ 4~~$ \\
     &                 & 0.081 & 0.005 & $ 8~~$ \\
     &                 & 0.084 & 0.008 & $10~~$ \\
     &                 & 0.075 & 0.008 & $37^p$ \\
     &                 & 0.063 & 0.006 & $ 7^p$ \\
     &                 & 0.070 & 0.002 & $99~~$ \\
   8 & Flora           & 0.043 & 0.005 & $18~~$ \\
     &                 & 0.033 & 0.004 & $ 8~~$ \\
     &                 & 0.037 & 0.007 & $10~~$ \\
     &                 & 0.022 & 0.004 & $37^p$ \\
     &                 & 0.034 & 0.004 & $ 7^p$ \\
     &                 & 0.029 & 0.002 & $99~~$ \\
   9 & Metis           & 0.103 & 0.024 & $ 4~~$ \\
     &                 & 0.057 & 0.014 & $18~~$ \\
     &                 & 0.012 & 0.010 & $ 7~~$ \\
     &                 & 0.057 & 0.011 & $ 8~~$ \\
     &                 & 0.045 & 0.007 & $10~~$ \\
     &                 & 0.035 & 0.007 & $37^p$ \\
     &                 & 0.042 & 0.007 & $ 7^p$ \\
     &                 & 0.025 & 0.002 & $99~~$ \\
  10 & Hygieia         & 0.557 & 0.070 & $17~~$ \\
     &                 & 0.501 & 0.041 & $14~~$ \\
     &                 & 0.25  & 0.04  & $19~~$ \\
     &                 & 0.454 & 0.015 & $ 4~~$ \\
     &                 & 0.436 & 0.007 & $ 8~~$ \\
     &                 & 0.413 & 0.015 & $10~~$ \\
     &                 & 0.528 & 0.042 & $37^p$ \\
     &                 & 0.440 & 0.032 & $ 7^p$ \\
     &                 & 0.410 & 0.004 & $99~~$ \\
  11 & Parthenope      & 0.026 & 0.001 & $15~~$ \\
     &                 & 0.032 & 0.001 & $ 4~~$ \\
     &                 & 0.029 & 0.001 & $10~~$ \\
     &                 &  0.02 &  0.00 & $ 9~~$ \\
  13 & Egeria          & 0.082 & 0.016 & $18~~$ \\
     &                 & 0.080 & 0.022 & $ 8~~$ \\
     &                 & 0.031 & 0.016 & $10~~$ \\
     &                 & 0.062 & 0.017 & $37^p$ \\
     &                 & 0.047 & 0.008 & $ 7^p$ \\
     &                 & 0.047 & 0.004 & $99~~$ \\
  14 & Irene           & 0.041 & 0.007 & $18~~$ \\
     &                 & 0.035 & 0.008 & $ 8~~$ \\
     &                 & 0.010 & 0.010 & $10~~$ \\
     &                 & 0.033 & 0.007 & $37^p$ \\
     &                 & 0.028 & 0.003 & $99~~$ \\
  15 & Eunomia         & 0.126 & 0.030 & $17~~$ \\
     &                 & 0.106 & 0.016 & $14~~$ \\
     &                 & 0.164 & 0.006 & $21~~$ \\
     &                 & 0.08  & 0.03  & $19~~$ \\
     &                 & 0.168 & 0.008 & $ 4~~$ \\
     &                 & 0.162 & 0.005 & $22~~$ \\
     &                 & 0.160 & 0.002 & $ 8~~$ \\
     &                 & 0.162 & 0.004 & $10~~$ \\
     &                 & 0.131 & 0.014 & $37^p$ \\
     &                 & 0.158 & 0.010 & $ 7^p$ \\
     &                 & 0.146 & 0.002 & $99~~$ \\
  16 & Psyche          & 0.087 & 0.026 & $23~~$ \\
     &                 & 1.27  & 0.18  & $24~~$ \\
     &                 & 0.338 & 0.028 & $25~~$ \\
     &                 & 0.134 & 0.022 & $14~~$ \\
     &                 & 0.40  & 0.14  & $19~~$ \\
     &                 & 0.129 & 0.017 & $ 4~~$ \\
     &                 & 0.114 & 0.004 & $ 8~~$ \\
     &                 & 0.123 & 0.008 & $10~~$ \\
     &                 & 0.089 & 0.021 & $67^p$ \\
     &                 & 0.126 & 0.022 & $ 7^p$ \\
     &                 & 0.117 & 0.002 & $99~~$ \\
  17 & Thetis          & 0.006 & 0.001 & $ 4~~$ \\
     &                 & 0.007 & 0.000 & $ 8~~$ \\
     &                 & 0.037 & 0.009 & $ 7^p$ \\
     &                 & 0.004 & 0.000 & $99~~$ \\
  18 & Melpomene       & 0.015 & 0.005 & $18~~$ \\
     &                 & 0.025 & 0.010 & $10~~$ \\
     &                 & 0.020 & 0.005 & $37^p$ \\
     &                 & 0.027 & 0.003 & $99~~$ \\
  19 & Fortuna         & 0.054 & 0.008 & $ 4~~$ \\
     &                 & 0.064 & 0.003 & $18~~$ \\
     &                 & 0.042 & 0.004 & $ 8~~$ \\
     &                 & 0.051 & 0.005 & $10~~$ \\
     &                 & 0.039 & 0.005 & $37^p$ \\
     &                 & 0.049 & 0.005 & $ 7^p$ \\
     &                 & 0.045 & 0.001 & $99~~$ \\
  20 & Massalia        & 0.024 & 0.004 & $26~~$ \\
     &                 & 0.029 & 0.004 & $18~~$ \\
     &                 & 0.017 & 0.004 & $ 8~~$ \\
     &                 & 0.030 & 0.006 & $10~~$ \\
     &                 & 0.029 & 0.011 & $ 7^p$ \\
     &                 & 0.024 & 0.001 & $99~~$ \\
  22 & Kalliope        & 0.037 & 0.002 & $27~~$ \\
     &                 & 0.085 & 0.028 & $14~~$ \\
     &                 & 0.041 & 0.001 & $29^s$ \\
     &                 & 0.041 & 0.001 & $28^s$ \\
     &                 & 0.066 & 0.019 & $10~~$ \\
     &                 & 0.084 & 0.007 & $ 7^p$ \\
     &                 & 0.024 & 0.004 & $99~~$ \\
  23 & Thalia          & 0.010 &       & $ 5^p$ \\
     &                 & 0.015 & 0.003 & $ 7^p$ \\
     &                 & 0.023 & 0.003 & $99~~$ \\
  24 & Themis          & 0.057 & 0.022 & $ 4~~$ \\
     &                 & 0.028 & 0.019 & $ 7^p$ \\
     &                 & 0.031 & 0.009 & $10~~$ \\
     &                 & 0.051 & 0.003 & $99~~$ \\
  27 & Euterpe         & 0.011 & 0.007 & $10~~$ \\
     &                 & 0.010 & 0.002 & $99~~$ \\
  28 & Bellona         & 0.07  & 0.02  & $19~~$ \\
     &                 & 0.008 & 0.003 & $99~~$ \\
  29 & Amphitrite      & 0.077 & 0.012 & $14~~$ \\
     &                 & 0.100 & 0.035 & $ 4~~$ \\
     &                 & 0.076 & 0.003 & $ 8~~$ \\
     &                 & 0.056 & 0.008 & $10~~$ \\
     &                 & 0.052 & 0.012 & $37^p$ \\
     &                 & 0.072 & 0.010 & $ 7^p$ \\
     &                 & 0.067 & 0.002 & $99~~$ \\
  31 & Euphrosyne      & 0.094 & 0.052 & $14~~$ \\
     &                 & 0.313 & 0.059 & $18~~$ \\
     &                 & 0.292 & 0.100 & $ 8~~$ \\
     &                 & 0.135 & 0.046 & $10~~$ \\
     &                 & 0.110 & 0.037 & $37^p$ \\
     &                 & 0.132 & 0.020 & $ 7^p$ \\
     &                 & 0.087 & 0.007 & $99~~$ \\
  39 & Laetitia        & 0.088 & 0.005 & $ 7^p$ \\
     &                 & 0.028 & 0.007 & $ 8~~$ \\
     &                 & 0.022 & 0.013 & $10~~$ \\
     &                 & 0.022 & 0.003 & $99~~$ \\
  41 & Daphne          & 0.092 & 0.026 & $ 7^p$ \\
     &                 & 0.024 & 0.028 & $10~~$ \\
     &                 & 0.039 & 0.009 & $37^p$ \\
     &                 & 0.047 & 0.007 & $99~~$ \\
  45 & Eugenia         & 0.030 & 0.001 & $30~~$ \\
     &                 & 0.09  & 0.03  & $19~~$ \\
     &                 & 0.029 & 0.001 & $29^s$ \\
     &                 & 0.038 & 0.008 & $10~~$ \\
     &                 & 0.032 & 0.002 & $99~~$ \\
  46 & Hestia          & 0.109 & 0.068 & $13~~$ \\
     &                 & 0.035 & 0.007 & $ 7^p$ \\
     &                 & 0.005 & 0.002 & $99~~$ \\
  48 & Doris           & 0.061 & 0.030 & $14~~$ \\
     &                 & 0.201 & 0.150 & $ 7^p$ \\
     &                 & 0.019 & 0.019 & $10~~$ \\
     &                 & 0.039 & 0.003 & $99~~$ \\
  49 & Pales           & 0.014 & 0.003 & $18~~$ \\
     &                 & 0.027 & 0.003 & $99~~$ \\
  51 & Nemausa         & 0.017 & 0.008 & $10~~$ \\
     &                 & 0.000 & 0.000 & $ 7^p$ \\
     &                 & 0.014 & 0.002 & $99~~$ \\
  52 & Europa          & 0.261 & 0.088 & $17~~$ \\
     &                 & 0.127 & 0.025 & $14~~$ \\
     &                 & 0.42  & 0.11  & $19~~$ \\
     &                 & 0.098 & 0.022 & $ 4~~$ \\
     &                 & 0.083 & 0.008 & $18~~$ \\
     &                 & 0.114 & 0.008 & $ 8~~$ \\
     &                 & 0.140 & 0.016 & $10~~$ \\
     &                 & 0.167 & 0.041 & $37^p$ \\
     &                 & 0.107 & 0.026 & $ 7^p$ \\
     &                 & 0.135 & 0.003 & $99~~$ \\
  59 & Elpis           & 0.026 & 0.018 & $ 7^p$ \\
     &                 & 0.014 & 0.002 & $10~~$ \\
     &                 & 0.017 & 0.003 & $99~~$ \\
  65 & Cybele          & 0.058 & 0.015 & $14~~$ \\
     &                 & 0.076 & 0.018 & $ 4~~$ \\
     &                 & 0.089 & 0.006 & $18~~$ \\
     &                 & 0.072 & 0.043 & $ 7^p$ \\
     &                 & 0.053 & 0.010 & $ 8~~$ \\
     &                 & 0.077 & 0.014 & $10~~$ \\
     &                 & 0.089 & 0.004 & $99~~$ \\
  68 & Leto            & 0.015 & 0.014 & $10~~$ \\
     &                 & 0.021 & 0.003 & $99~~$ \\
  69 & Hesperia        & 0.033 & 0.021 & $10~~$ \\
     &                 & 0.047 & 0.005 & $99~~$ \\
  85 & Io              & 0.018 & 0.013 & $10~~$ \\
     &                 & 0.022 & 0.008 & $ 7^p$ \\
     &                 & 0.012 & 0.003 & $99~~$ \\
  87 & Sylvia          & 0.075 & 0.000 & $30^s$ \\
     &                 & 0.26  & 0.11  & $19~~$ \\
     &                 & 0.059 & 0.053 & $10~~$ \\
     &                 & 0.075 & 0.001 & $36^s$ \\
     &                 & 0.108 & 0.010 & $99~~$ \\
  88 & Thisbe          & 0.074 & 0.013 & $17~~$ \\
     &                 & 0.059 & 0.012 & $14~~$ \\
     &                 & 0.057 & 0.018 & $ 4~~$ \\
     &                 & 0.053 & 0.009 & $18~~$ \\
     &                 & 0.092 & 0.006 & $ 8~~$ \\
     &                 & 0.038 & 0.011 & $10~~$ \\
     &                 & 0.057 & 0.017 & $37^p$ \\
     &                 & 0.071 & 0.014 & $ 7^p$ \\
     &                 & 0.051 & 0.002 & $99~~$ \\
  89 & Julia           & 0.042 & 0.021 & $ 7^p$ \\
     &                 & 0.043 & 0.003 & $99~~$ \\
  94 & Aurora          & 0.150 & 0.034 & $ 7^p$ \\
     &                 & 0.009 & 0.003 & $99~~$ \\
  96 & Aegle           & 0.040 & 0.014 & $37^p$ \\
     &                 & 0.057 & 0.009 & $99~~$ \\
 106 & Dione           & 0.018 & 0.013 & $10~~$ \\
     &                 & 0.039 & 0.004 & $ 7^p$ \\
     &                 & 0.018 & 0.003 & $99~~$ \\
 107 & Camilla         & 0.056 & 0.002 & $29^s$ \\
     &                 & 0.088 & 0.044 & $10~~$ \\
     &                 & 0.034 & 0.015 & $ 7^p$ \\
     &                 & 0.056 & 0.009 & $99~~$ \\
 117 & Lomia           & 0.087 & 0.001 & $ 7^p$ \\
     &                 & 0.034 & 0.018 & $10~~$ \\
     &                 & 0.017 & 0.004 & $99~~$ \\
 121 & Hermione        & 0.047 & 0.008 & $23~~$ \\
     &                 & 0.027 & 0.002 & $31^s$ \\
     &                 & 0.024 & 0.001 & $32^s$ \\
     &                 & 0.023 & 0.011 & $10~~$ \\
     &                 & 0.016 & 0.002 & $99~~$ \\
 128 & Nemesis         & 0.042 & 0.011 & $10~~$ \\
     &                 & 0.069 & 0.016 & $ 7^p$ \\
     &                 & 0.042 & 0.003 & $99~~$ \\
 129 & Antigone        & 0.043 & 0.018 & $ 7^p$ \\
     &                 & 0.024 & 0.028 & $10~~$ \\
     &                 & 0.032 & 0.004 & $99~~$ \\
 130 & Elektra         & 0.033 & 0.002 & $27^s$ \\
     &                 & 0.112 & 0.080 & $ 7^p$ \\
     &                 & 0.035 & 0.032 & $10~~$ \\
     &                 & 0.070 & 0.008 & $99~~$ \\
 139 & Juewa           & 0.030 & 0.016 & $10~~$ \\
     &                 & 0.021 & 0.009 & $ 7^p$ \\
     &                 & 0.030 & 0.003 & $99~~$ \\
 144 & Vibilia         & 0.046 & 0.030 & $ 7^p$ \\
     &                 & 0.025 & 0.009 & $10~~$ \\
     &                 & 0.017 & 0.002 & $99~~$ \\
 145 & Adeona          & 0.011 & 0.015 & $10~~$ \\
     &                 & 0.014 & 0.003 & $99~~$ \\
 150 & Nuwa            & 0.091 & 0.004 & $ 7^p$ \\
     &                 & 0.010 & 0.014 & $10~~$ \\
     &                 & 0.017 & 0.003 & $99~~$ \\
 154 & Bertha          & 0.064 & 0.048 & $10~~$ \\
     &                 & 0.081 & 0.008 & $99~~$ \\
 173 & Ino             & 0.067 & 0.015 & $ 7^p$ \\
     &                 & 0.016 & 0.003 & $99~~$ \\
 187 & Lamberta        & 0.025 & 0.011 & $ 7^p$ \\
     &                 & 0.011 & 0.003 & $99~~$ \\
 194 & Prokne          & 0.056 & 0.006 & $ 7^p$ \\
     &                 & 0.023 & 0.003 & $99~~$ \\
 196 & Philomela       & 0.017 & 0.008 & $10~~$ \\
     &                 & 0.027 & 0.004 & $99~~$ \\
 211 & Isolda          & 0.038 & 0.007 & $ 7^p$ \\
     &                 & 0.012 & 0.009 & $10~~$ \\
     &                 & 0.015 & 0.003 & $99~~$ \\
 212 & Medea           & 0.066 & 0.005 & $ 7^p$ \\
     &                 & 0.011 & 0.001 & $99~~$ \\
 216 & Kleopatra       & 0.006 & 0.005 & $ 7^p$ \\
     &                 & 0.038 & 0.011 & $10~~$ \\
     &                 & 0.026 & 0.004 & $99~~$ \\
 238 & Hypatia         & 0.031 & 0.016 & $10~~$ \\
     &                 & 0.015 & 0.003 & $99~~$ \\
 259 & Aletheia        & 0.037 & 0.023 & $10~~$ \\
     &                 & 0.065 & 0.005 & $99~~$ \\
 268 & Adorea          & 0.008 & 0.007 & $ 7^p$ \\
     &                 & 0.016 & 0.002 & $99~~$ \\
 324 & Bamberga        & 0.229 & 0.038 & $14~~$ \\
     &                 & 0.054 & 0.010 & $10~~$ \\
     &                 & 0.051 & 0.004 & $37^p$ \\
     &                 & 0.048 & 0.004 & $ 7^p$ \\
     &                 & 0.044 & 0.002 & $99~~$ \\
 328 & Gudrun          & 0.009 & 0.026 & $10~~$ \\
     &                 & 0.029 & 0.007 & $99~~$ \\
 349 & Dembowska       & 0.018 & 0.013 & $10~~$ \\
     &                 & 0.014 & 0.003 & $99~~$ \\
 354 & Eleonora        & 0.060 & 0.025 & $10~~$ \\
     &                 & 0.089 & 0.006 & $99~~$ \\
 375 & Ursula          & 0.037 & 0.020 & $10~~$ \\
     &                 & 0.059 & 0.006 & $99~~$ \\
 409 & Aspasia         & 0.017 &       & $ 5^p$ \\
     &                 & 0.062 & 0.016 & $10~~$ \\
     &                 & 0.000 & 0.000 & $ 7^p$ \\
     &                 & 0.027 & 0.003 & $99~~$ \\
 423 & Diotima         & 0.034 & 0.021 & $10~~$ \\
     &                 & 0.048 & 0.004 & $99~~$ \\
 451 & Patientia       & 0.102 & 0.034 & $14~~$ \\
     &                 & 0.056 & 0.032 & $10~~$ \\
     &                 & 0.150 & 0.036 & $ 7^p$ \\
     &                 & 0.068 & 0.005 & $99~~$ \\
 488 & Kreusa          & 0.011 & 0.034 & $10~~$ \\
     &                 & 0.037 & 0.005 & $99~~$ \\
 511 & Davida          & 0.334 & 0.028 & $17~~$ \\
     &                 & 0.240 & 0.024 & $14~~$ \\
     &                 & 0.298 & 0.030 & $ 4~~$ \\
     &                 & 0.220 & 0.010 & $18~~$ \\
     &                 & 0.190 & 0.010 & $ 8~~$ \\
     &                 & 0.131 & 0.030 & $10~~$ \\
     &                 & 0.148 & 0.042 & $37^p$ \\
     &                 & 0.183 & 0.029 & $ 7^p$ \\
     &                 & 0.151 & 0.005 & $99~~$ \\
 532 & Herculina       & 0.168 & 0.028 & $14~~$ \\
     &                 & 0.088 & 0.022 & $10~~$ \\
     &                 & 0.065 & 0.014 & $37^p$ \\
     &                 & 0.116 & 0.001 & $  ~~$ \\
     &                 & 0.104 & 0.004 & $99~~$ \\
 690 & Wratislavia     & 0.064 & 0.001 & $ 7^p$ \\
     &                 & 0.011 & 0.005 & $99~~$ \\
 702 & Alauda          & 0.031 & 0.002 & $35^s$ \\
     &                 & 0.069 & 0.056 & $ 7^p$ \\
     &                 & 0.066 & 0.010 & $99~~$ \\
 704 & Interamnia      & 0.352 & 0.093 & $17~~$ \\
     &                 & 0.081 & 0.042 & $14~~$ \\
     &                 & 0.57  & 0.16  & $19~~$ \\
     &                 & 0.358 & 0.042 & $ 4~~$ \\
     &                 & 0.186 & 0.011 & $18~~$ \\
     &                 & 0.157 & 0.026 & $10~~$ \\
     &                 & 0.198 & 0.035 & $37^p$ \\
     &                 & 0.192 & 0.024 & $ 7^p$ \\
     &                 & 0.137 & 0.006 & $99~~$ \\
 790 & Pretoria        & 0.022 & 0.024 & $10~~$ \\
     &                 & 0.049 & 0.009 & $99~~$ \\
 804 & Hispania        & 0.047 & 0.019 & $ 4~~$ \\
     &                 & 0.020 & 0.004 & $18~~$ \\
     &                 & 0.018 & 0.004 & $ 8~~$ \\
     &                 & 0.025 & 0.018 & $ 7^p$ \\
     &                 & 0.028 & 0.014 & $10~~$ \\
     &                 & 0.014 & 0.004 & $99~~$ \\
\end{longtable}
\end{longtab}
\tablebib{
  (1) Viateau \& Rapaport		(\cite{Viat1998});
  (2) Michalak				(\cite{Mich2000});
  (3) Kova\u{c}evi\'c \& Kuzmanoski	(\cite{Kova2007});
  (4) Baer \& Chesley			(\cite{Baer2008a});
  (5) Folkner et al.			(\cite{Folk2008});
  (6) Pitjeva				(\cite{Pitj2009});
  (7) Fienga et al.			(\cite{Fien2013});
  (8) Baer et al.			(\cite{Baer2011});
  (9) Konopliv et al.			(\cite{Kono2011});
 (10) Zielenbach			(\cite{Ziel2011});
 (11) Goffin				(\cite{Goff2001});
 (13) Bang \& Bec-Borsenberger		(\cite{Bang1997});
 (14) Kochetova				(\cite{Koch2004});
 (15) Viateau \& Rapaport		(\cite{Viat2001});
 (16) Kova\u{c}evi\'c			(\cite{Kova2005});
 (17) Michalak				(\cite{Mich2001});
 (18) Baer et al.			(\cite{Baer2008b});
 (19) Ivantsov				(\cite{Ivan2008});
 (20) Kuzmanoski et al.			(\cite{Kuzm2010});
 (21) Vitagliano \& Stoss		(\cite{Vita2006});
 (22) Zielenbach			(\cite{Ziel2010});
 (23) Viateau				(\cite{Viat2000});
 (24) Krasinsky et al.			(\cite{Krasy2001};
 (25) Kuzmanoski \& Kova\u{c}evi\'c	(\cite{Kuzm2002});
 (26) Bange				(\cite{Bang1998});
 (27) Marchis et al.			(\cite{Marc2008a});
 (28) Descamps et al.			(\cite{Desc2008});
 (29) Marchis et al.			(\cite{Marc2008b});
 (30) Marchis et al.			(\cite{Marc2005a});
 (31) Marchis et al.			(\cite{Marc2005b});
 (32) Descamps et al.			(\cite{Desc2009});
 (33) Merline et al.			(\cite{Merl2002});
 (34) Russell et al.			(\cite{Russ2012});
 (35) Rojo \& Margot			(\cite{Rojo2010});
 (36) Fang et al.			(\cite{Fang2012});
 (37) Kuchynka \& Folkner		(\cite{Kuch2013});
 (99) This paper.}
}

\end{document}